%% file: main.tex
\newif\ifdraft \drafttrue \draftfalse
\newif\ifincludeappendix \includeappendixtrue 
\documentclass[a4paper, 12pt, twoside]{article}
\usepackage[utf8]{inputenc}
\usepackage{tgpagella,eulervm}
\usepackage[OT1]{fontenc}
\usepackage[%
  ,sorting=anyt%
  ,minalphanames=2%
  ,maxalphanames=3%
  ,maxbibnames=100%
  ,style=alphabetic%
]{biblatex}
\DeclareLabelalphaTemplate{
    \labelelement{
        \field[final]{shorthand}
        \field{label}
        \field[strwidth=1,strside=left]{labelname}
    }
    \labelelement{
        \literal{\hspace*{.1em}}
    }
    \labelelement{
     \field[strwidth=2,strside=right]{year}
    }
}

\bibliography{bibliography.bib}

\usepackage{amsthm,amsmath,amssymb}
\usepackage{xargs,calc}
\usepackage{stmaryrd}
\usepackage[head=6mm,margin=28mm, foot=16mm, top=30mm, bottom=36mm]{geometry}
\usepackage{xcolor}
\usepackage{graphicx}
\usepackage{subfig}
\usepackage[blocks]{authblk}
\usepackage{xspace,adjustbox}
\usepackage[inline]{enumitem}
\usepackage{etoolbox}
\setlist{parsep=0pt, itemsep={0.2\thmspace}, topsep={0.3\thmspace}}
\setlist[enumerate]{label=\enumstyle{\alph{*}}}
\newlist{enuminline}{enumerate*}{2}
\setlist[enuminline]{label={\hspace{.5ex plus .3ex minus .2ex}\rm\arabic{*})},itemjoin*={{ }and{ }}}
\setlist[trivlist]{parsep=0pt,partopsep=0pt,topsep=\thmspace}
\newlist{subthm}{enumerate}{3}
\setlist[subthm]{label=\enumstyle{\alph{*}},ref=\thethm\hspace*{1pt}\alph{*}}

\newcommand{\enumstyle}[1]{{\rm(#1)}}

\usepackage{fancyhdr}
\ifdraft
\else
\fi

\ifdraft
  \usepackage[mathlines,right]{lineno}
  \linenumbers
  \usepackage[notref,notcite]{showkeys}
  \renewcommand*\showkeyslabelformat[1]{\scriptsize\normalfont\ttfamily #1}
\else
  \newcommand*\showkeyslabelformat[1]{}
\fi

\apptocmd\normalsize{%
 \abovedisplayskip=6pt plus 1pt minus 0.5pt
 \abovedisplayshortskip=0pt plus 2pt
 \belowdisplayskip=6pt plus 1pt minus 0.5pt
 \belowdisplayshortskip=0pt plus 2pt
}{}{}

\usepackage{multicol}
\DeclareMathSymbol{\in}{\mathbin}{symbols}{"32}
\newcommand{\noneqspacing}{
  \thickmuskip=5mu plus 3mu minus 2mu
  \medmuskip=3mu plus 2mu minus 2mu
}
\noneqspacing
\newcommand{\eqspacing}{
  \thickmuskip=12mu plus 3mu minus 3mu
  \medmuskip=5mu plus 3mu minus 2mu
}%
\AtBeginEnvironment{gather*}{\eqspacing}%
\AtBeginEnvironment{gather}{\eqspacing}%
\AtBeginEnvironment{equation}{\eqspacing}%
\AtBeginEnvironment{equation*}{\eqspacing}%
\AtBeginEnvironment{align}{\eqspacing}%
\AtBeginEnvironment{align*}{\eqspacing}%
\AtBeginEnvironment{multline}{\eqspacing}%
\AtBeginEnvironment{multline*}{\eqspacing}%

\ifdraft
\fi

\usepackage[beginComment={//~},beginLComment={//~},endLComment={}]{algpseudocodex}
\tikzset{algpxIndentLine/.style={dotted}}

\newcounter{lastalgoline}
\newenvironment{vmalgorithm}%
{\begin{algorithmic}[1]\setcounter{ALG@line}{\value{lastalgoline}}}
{\setcounter{lastalgoline}{\value{ALG@line}}\end{algorithmic}}

\usepackage[scaled=0.8]{FiraMono}
\algrenewcommand\alglinenumber[1]{\makebox[1em][r]{\footnotesize #1.}}
\algrenewcommand\algorithmicdo{}
\algrenewcommand\algorithmicthen{}

\algrenewcommand\algorithmicforall{\textbf{for each}}

\input{aux/macros}
\input{aux/vmthm}
\newcommand{\lex}{\mathrel{\leq_{LEX}}}
\newcommand{\rad}{\mathrel{\leq_{RAD}}}
\usepackage{tikz}
\usetikzlibrary{arrows.meta,calc,shapes}
\input{aux/tikzstyle.tex}

\usepackage{svg,hyperref}

\newcommand{\thetitle}{Enumerating regular languages in radix order}
\newcommand{\thesubtitle}{Revisiting the Ackerman-Shallit algorithm}
\newcommand{\thefulltitle}{\thetitle\ifdefempty{\thesubtitle}{}{: \thesubtitle}}
\newcommand{\theshorttitle}{}
\title{\thetitle\ifdefempty{\thesubtitle}{}{\\[.25em] \Large \thesubtitle}}

\author{Nadime Francis}
\affil{LIGM, Université Gustave Eiffel, CNRS\\
   \orcidfull{https://orcid.org/0009-0009-4531-7435}\\
   \mel{nadime.francis}{univ-eiffel.fr}%
}

\author{Victor Marsault}
\affil{LIGM, Université Gustave Eiffel, CNRS\\
   \orcidfull{https://orcid.org/0000-0002-2325-6004}\\
   \mel{victor.marsault}{univ-eiffel.fr}%
}

\date{\isotoday}

\hypersetup{%
  hidelinks,
  pdftitle={\protect\thefulltitle},
}

\begin{document}

\maketitle

\begin{abstract}
    We consider the problem of enumerating a regular language~$L$ in radix order, or more precisely, the equivalent problem of enumerating all words in~$L$ of a given length~$\ell$ in lexicographic order.
    Ackerman and Shallit gave in 2009 the principles of an efficient solution to this problem, but they did not use the enumeration complexity framework for their analysis.
    We adapt their work into an explicit algorithm that fits this framework.
\end{abstract}

\section{Introduction}

We consider the problem of enumerating all the words in a given regular language~$L$ in radix order~$\rad$ (also called shortlex order), that is by increasing length, and for words of the same length, in lexicographic order.
This is done by enumerating, for each $\ell$, the set of words of $L$ of length $\ell$ (called a \emph{cross-section}) in lexicographic order.
Using a prior idea of Mäkinen \cite{Makinen1997}, 
Ackerman and Shallit \cite{AckermanShallit2009} give an efficient algorithm for the \problemfont{Cross-section} problem, defined below.

\begin{problem}{Cross-section}
    \item[Input:] A nondeterministic automaton~$\Ac=\aut{\Sigma,Q,I,\Delta,F}$ and an integer $\ell$.
    \item[Output:] Enumerate all words of length $\ell$ in~$L(\Ac)$ in lexicographic order.
\end{problem}

In \cite{AckermanShallit2009}, the algorithm and its associated complexity analysis are given for producing the full output. However, this does not satisfy the needs of some scenarios in which the amount of words to enumerate might be unreasonably large.
Indeed, Martens and Trautner \cite{MartensTrautner2018} quite recently reduced a problem
coming from database theory to \problemfont{Cross-section}.
In database theory, enumeration complexity is the norm: the output is likely to be large and
one expects to be able to stream partial results to continue computations.


Martens and Trautner used only the fact that \problemfont{Cross-section} can be enumerated with polynomial delay and preprocessing, something that is clear from \cite{AckermanShallit2009}.
However, fine-grained enumeration complexity cannot be easily derived from the algorithm given in \cite{AckermanShallit2009}.
In this note, we give a version of the Ackerman-Shallit algorithm that fits the enumeration complexity framework
(see for instance chapter~3 of \cite{Strozecki2021}). 
This leads to the following statement.

\begin{theorem}[\citeauthoryear{AckermanShallit2009}]
    \label{t:ackerman-shallit}
    \problemfont{Cross-section} can be enumerated with $\bigo{\card{\Sigma} \times \card{Q} + \ell\times\card[2]{Q}+ \ell\times\card{\Delta}}$ preprocessing and $\bigo{\ell\times\card{\Delta}}$ delay.
    
    Depending on the data structure for implementing $\Ac$, the $\card{\Sigma}\times\card{Q}$ preprocessing term can be avoided. 
\end{theorem}

The general principle of the algorithm comes from the work of Mäkinen \cite{Makinen1997}. In our case, it is adapted as follows.
We first find the least word of length $\ell$ in $L(\Ac)$. Then, from any word $w\in L(\Ac)$ of length $\ell$, we look for the least word~$w'$ in $L(\Ac)$ such that~$w<_{LEX} w'$ and $\len(w) = \len(w')$. Recall that these last two conditions imply that there exist three words $u,v,v'$ and two letters $a, a'$ such that:
\begin{equation}
    w = uav \quad;\qquad w' = ua'v' \quad;\qquad \len(v) = \len(v') \quad;\qquad a < a'
\end{equation}
Finding $w'$ amounts to finding the least word $a'v'$ in $\rad$ for which $ua'v'\in L(\Ac)$ and Equation~(\theequation{}) holds.
In order to make this search as efficient as possible during the enumeration phase, the preprocessing phase precomputes data structures that allow retrieving the least words that are accepted by $\Ac$ starting from each possible state $q$, and for each possible length $k \leq \ell$.

\medskip

The correctness of the algorithm immediately comes from the arguments in \cite{Makinen1997} or \cite{AckermanShallit2009}, as we follow the existing principles very closely. The main change consists in carefully designing the precomputed structures in order to have a low and consistent delay.

\subsection*{Outline}
First, Section~\ref{s:prelim} gives the definition we use for automata and the assumption on its
representation in memory.
Then, the preprocessing phase of the algorithm is explained in Section~\ref{s:preprocessing},
as well as its complexity.
Similarly, Section~\ref{s:enumeration} explains the enumeration phase and its complexity.
The complete pseudo-code of the algorithm is given in pages~\pageref{preproc:l loop start} and~\pageref{nextword:clear delta}.

\section{Automata: definition and memory representation}
\label{s:prelim}

\begin{definition}
    A (nondeterministic finite) \emph{automaton} is a 5-tuple $\Ac=\aut{\Sigma,Q,I,\Delta,F}$ where $\Sigma$ is a finite set of \emph{symbols} or \emph{letters}, and is called the \emph{alphabet} of~$\Ac$;~$Q$ is a finite set of \emph{states};~$I\subseteq Q$ is called the set of \emph{initial} states; $\Delta\subseteq{Q\times \Sigma\times Q}$ is the set of \emph{transitions} and~$F\subseteq Q$ is the set of \emph{final} states.
\end{definition}

Since we consider fine-grained complexity, we need to explicit several assumptions about the memory representation of automata.
The data structure underlying the input automaton $A=\aut{\Sigma,Q,I,\Delta,F}$ is assumed to satisfy the following properties.
\begin{enumerate}[align=left,label=(H\arabic{*}),leftmargin=*]
    \item $Q$ is equal to $\{0,\ldots,\card{Q}-1\}$.
    \item \label{hypothesis:delta(q,a)}
    For each state $q$ and letter $a$, one may access the state list $\Delta(q,a)$ in $\bigo{1}$ time; this list contains all states $q'$ such that $(q,a,q')\in\Delta$, in no particular order but without duplicates.
    \item \label{hypothesis:delta(q)}
    For each $q$, one may access the list $\Delta(q)$ in $\bigo{1}$ time.
     More precisely, $\Delta(q)$ is the list of the pairs $(a,\Delta(q,a))$ such that $\Delta(q,a)\neq 0$, ordered by increasing $a$.
    In particular, when $q$ has no outgoing transition for a given $a$, no pair $(a,\cdot)$ will be present in the list.
        
    \item The sizes of the components of~$\Ac$ satisfy: $\card{\Sigma}\leq\card{\Delta}$ and $\card{Q}\leq\card{\Delta}$.
\end{enumerate}

\medskip

\noindent Moreover, we assume the following hypotheses for data structures.
\begin{enumerate}[align=left,label=(H\arabic{*}),leftmargin=*,resume]
    \item \label{hypothesis:state set}
    We use a data structure for \emph{state sets}, that is subsets of~$Q$, that provide the following operations.
        \begin{itemize}
            \item Creating a new empty state set uses $\bigo{\card{Q}}$ time.
            \item Inserting a state may be done in $\bigo{1}$.
            \item Iterating through a state set~$S$ (in no particular order) may be done in $\bigo{\card{S}}$.
            \item Clearing a state set~$S$ may be done in $\bigo{\card{S}}$.
        \end{itemize}
        This may be achieved by using a sparse array, or more simply by maintaining a boolean array of length~$\card{Q}$ 
        and a linked list of the elements that are present.
    \item \label{hypothesis:maps} For maps ($\minarrow$ and $\comp$) we use fully initialised array/matrices. Hence creation takes $\bigo{D}$ time, where~$D$ is the domain of the map, and modification/access uses $\bigo{1}$ time.
\end{enumerate}

\begin{remark}\label{remark:delta(q)} 
If \ref{hypothesis:delta(q)} is not true, precomputing a map realising $q\mapsto\Delta(q)$ using \ref{hypothesis:delta(q,a)} uses $\bigo{\card{\Delta}+\card{\Sigma}\times\card{Q}}$ time. Since $\card{\Sigma}\times\card{Q}$ is in general incomparable with $\card{\Delta}$ or $\card[2]{Q}$,
and since \ref{hypothesis:delta(q)} is nonstandard, we added it to the preprocessing time complexity in Theorem~\ref{t:ackerman-shallit}.
\end{remark}

\begin{figure}[p]
\begin{vmalgorithm}
\Function{Main}{Automaton $\Ac=\aut{\Sigma,Q,I,\Delta,F}$, length~$\ell$}
    \State $(\minarrow, \comp) \gets \Call{Preproc}{\Ac,\ell}$
    \State $w \gets \Call{MinWord}{\ell, I, \minarrow, \comp}$
    \While{$w \neq \bot$}
        \State \textbf{output} $w$
        \State $S\gets\Call{BuildStack}{w, \Ac}$
        \State $w \gets \Call{NextWord}{w,\ell,\Ac,S, \minarrow, \comp}$
    \EndWhile
\EndFunction
\end{vmalgorithm}
\end{figure}

\begin{figure}[p]
\begin{vmalgorithm}
\Function{Preproc}{Automaton $\Ac=\aut{\Sigma,Q,I,\Delta,F}$, length $\ell$}
    \State $\minarrow \gets $ new Map $\set{0,...,\ell}\to Q \to (\Sigma\times Q) \uplus \set{\bot, \varepsilon}$, initialised with $\bot$
    \State $\comp \gets $ new Map $\set{0,...,\ell} \to Q\times Q \to \set{true, false}$, initialised with $false$
    \ForAll{$q\in F$}
        \State $\minarrow[0][q] \gets \varepsilon$
        \ForAll{$q'\in Q$}
            \State $\comp[0][q,q'] \gets true$
            
        \EndFor
    \EndFor
    \ForAll{$k$ from $1$ to $\ell$}\label{preproc:l loop start}
        \ForAll{$q\in Q$} \label{preproc:for each q}
            \LComment{\raggedright We compute $\minarrow[k][q]$ from $\minarrow[k-1]$ and $\comp[k-1]$}
            \ForAll{$(a,L) \in \Delta(q)$} 
                \State $q_{min} \gets L[0]$\Comment{$L$ cannot be empty (see \ref{hypothesis:delta(q)})}
                \ForAll{$q'\in L$}
                    \If{$\comp[k-1][q',q_{min}]$}
                        \State $q_{min} \gets q'$
                    \EndIf
                \EndFor
                \If{$\minarrow[k-1][q_{min}] \neq \bot$}
                    \State $\minarrow[k][q] \gets (a,q_{min})$
                    \State \textbf{break}
                \LComment{Recall that $\Delta(q)$ is sorted by increasing $a$ (see \ref{hypothesis:delta(q)})}
                \LComment{Hence $a$ is the least letter where a suitable $q_{min}$ exists}
                \EndIf
                \label{preproc:end for each q}
            \EndFor
        \EndFor
        \ForAll{$(q,q') \in Q\times Q$}\label{preproc:for each (q,q')}
            \LComment{\raggedright We compute $\comp[k][q,q']$ from $\minarrow[k]$ and $\comp[k-1]$}
            \State $t \gets \minarrow[k][q]$
            \State $t' \gets \minarrow[k][q']$
            \If{$t \neq \bot$}
                \If{$t' = \bot$}
                    \State $\comp[k][q,q'] \gets true$
                \Else
                    \State $(a,p) \gets t$ \commentpar{$k>0$ hence $t=\varepsilon$ is impossible}                    
                    \State $(a',p') \gets t'$\commentpar{$k>0$ hence $t'=\varepsilon$ is impossible}
                    \If{ $a < a'$ \textbf{or} $(a = a'$ \textbf{and} $\comp[k-1][p,p'])$}
                        \State $\comp[k][q,q'] \gets true$
                    \EndIf
                \EndIf
            \EndIf
            \label{preproc:end for each (q,q')}
        \EndFor
    \EndFor
    \State \Return $(\minarrow, \comp)$
\EndFunction
\end{vmalgorithm}
\end{figure}

\section{Preprocessing}
\label{s:preprocessing}

The goal of the preprocessing phase is to compute two maps $\minarrow : \set{0,...,\ell} \to Q \to (\Sigma\times Q) \uplus \set{\bot, \varepsilon}$ and $\comp : \set{0,...,\ell} \to Q\times Q \to   \set{true, false}$, described below.
When it exists, we denote by $w_{k,q}$ the least word in $\rad$ among the words~$w$ of length~$k$ that are accepted from a given state~$q$.
\begin{itemize}
    \item For each~$k$, $\minarrow[k]$ generally maps a state $q$ to the first transition of some computation of~$w_{k,q}$ starting from~$q$.  It returns $\bot$ if $\Ac$ accepts no word of length~$k$ from $q$. It returns $\varepsilon$ if $\ell = 0$ and $q\in F$; indeed, if it exists, $w_{0,q}$ is equal to~$\varepsilon$.
    \item For each~$k$, $\comp[k]$ implements a pre-order~$\leq_{k}$ on~$Q$ defined by: 
    $q\leq_{k} q'$ if:
    \begin{itemize}[align=left,leftmargin=*]
        \item either $\Ac$ accepts a word of length $k$ from $q$ but not from $q'$;
        \item or $w_{k,q} \lex w_{k,q'}$.
    \end{itemize}
    Note that~$\leq_{k}$ is \textbf{not} antisymmetric, hence it is possible for both $\comp[k][q,q']$  and $\comp[k][q',q]$ to be equal to $true$ for different $q,q'$; it only means that $w_{k,q}$ and $w_{k,q'}$, as defined above, are equal.
    Note that~$\leq_{k}$ is \textbf{not} total, hence it is possible for both $\comp[k][q,q']$  and $\comp[k][q,q']$ to be equal to $false$;
    it means that~$\Ac$ accepts no word of length~$k$ from $q$ nor $q'$.
\end{itemize}

\medskip

\noindent Then, the complexity breaks down as follows.
\begin{itemize}
    \item Creating $\minarrow$ takes $\bigo{\ell\times\card{Q}}$ time and computing it (lines \ref{preproc:for each q}--\ref{preproc:end for each q}) takes $\bigo{\ell\times\card{\Delta}}$ time.
    \item Creating and computing $\comp$ both take $\bigo{\ell\times\card[2]{Q}}$ time.
\end{itemize}
The complete preprocessing uses  $\bigo{\ell\times\card{\Delta}+\ell\times\card[2]{Q}}$ time,
and requires an extra $\bigo{\card{\Sigma}\times\card{Q}}$ time in case hypothesis $\ref{hypothesis:delta(q)}$ is not met (see Remark~\ref{remark:delta(q)}).

\begin{figure}[p]
\begin{vmalgorithm}
\Function{MinWord}{Length $k$, State Set $P$, Map $\minarrow$, Map $\comp$}
    \LComment{First, using $\comp[k]$ we compute (one of) the states~$q$ in~$P$ such that $w$ is accepted from~$q$, where $w$ is the smallest word of length~$k$ accepted from some state in~$P$ }
    \If{each state $q\in P$ satisfies $\minarrow[k][q] = \bot$}
        \State \Return $\bot$ \Comment{No word of length~$k$ is accepted from any state in $P$}
    \EndIf 
    \State $q_{min} \gets P[0]$
    \ForAll{$q\in P$}
        \If{$\comp[k][q,q_{min}]$}
            \State $q_{min}\gets q$
        \EndIf
    \EndFor

    \LComment{Second, we follow the minimal transitions to reconstruct the smallest word accepted from $q$ in lexicographic order}
    \State $w\gets \varepsilon$
    \While{$\minarrow[k][q] \neq \varepsilon$}
        \State $(a,q) \gets \minarrow[k][q]$
        \State $k \gets k-1$
        \State $w \gets w\cdot a$
    \EndWhile
    \State \Return $w$
\EndFunction
\end{vmalgorithm}
\end{figure}

\begin{figure}[p]
\begin{vmalgorithm}
\Function{BuildStack}{Word $w$, Automaton $\Ac=\aut{\Sigma,Q,I,\Delta,F}$}
    \State $S \gets $ new empty Stack
    \State push $I$ on top of $S$
    \State $P \gets I$
    \ForAll{$a \in w$}
        \State $P \gets \Delta(P, a)$
        \State push $P$ on top of $S$
    \EndFor
    \State \Return $S$
\EndFunction
\end{vmalgorithm}
\end{figure}

\begin{figure}[p]
\begin{vmalgorithm}
\Function{NextWord}{Word $w$, Length $\ell$, Automaton $\Ac=\aut{\Sigma,Q,I,\Delta,F}$, Stack $S$, Map \minarrow, Map \comp}
    \State $P \gets$ new empty State Set
    \ForAll{$i$ from $\ell-1$ down to $0$}\label{nextword:i loop start}
        \State $cur \gets $ pop$(S)$
        \ForAll{$a \in \Sigma$ such that $a > w[i]$}
        \Comment{loop over $\Sigma$ from least to greatest symbol}
            \State add to $P$ all elements in $\Delta(cur,a)$ \label{nextword:delta}
            \State $w' \gets \Call{MinWord}{\ell-i,P,\minarrow,\comp}$\label{nextword:minword}
            \If{$w' \neq \bot$}
                \State \Return $w[0:i]\cdot a\cdot w'$
            \EndIf
            \State clear$(P)$ \label{nextword:clear delta}
        \EndFor
    \EndFor
    \State \Return $\bot$
\EndFunction
\end{vmalgorithm}
\end{figure}

\section{Enumeration}
\label{s:enumeration}

The enumeration phase starts from the minimal word that is accepted by $\Ac$ and then iteratively transforms it into the next words in lexicographic order, until no more accepted words can be produced. More precisely:

\begin{enumerate}
    \item The computation of the next word starts with the last output $w$.
    
    \item \functionfont{BuildStack} computes the run of~$w$ in~$\Ac$, and returns it as a stack~$S$ of state sets.
    For each $i$, $q\in S[i]$ if and only if $\Ac$ can reach $q$ from an initial state after reading the prefix of length $i$ of $w$, where $S[0]$ is the bottom of the stack and $S[\ell]$ the top of the stack.

    \item \functionfont{NextWord} computes the least word~$w'$ in $L(\Ac)$ such that~$w <_{RAD} w'$.
    It does so by attempting to replace longer and longer suffixes of $w$.
    For increasing~$i$ in~$\{0,\ldots,\ell-1\}$,
    factorise $w=uav$ with $\len(v)=i$ and check if there is any word in $L(\Ac)$ of the form $ua'v'$ with $a'> a$ and $\len(v')=i$.
    The algorithm is written in such a way
    that~$a'$ is as small as possible and uses \functionfont{MinWord}
    to find~$v'$, ensuring that $v'$ is also as small as possible.

    \item When the next word $w'$ has been found, it is output and we start over with $w\gets w'$. If no $w'$ can be found, the algorithm stops.
\end{enumerate}

\medskip

\noindent Here is the complexity analysis for the enumeration part of the algorithm.

\begin{itemize}
    \item The complexity of function \functionfont{MinWord} is different depending of its return value.
    When it returns $\bot$, it costs $\bigo{\card{P}}$.
    When it returns a word, it runs in $\bigo{\ell+\card{P}}$ time (in fact $\bigo{k+\card{P}}$) since access to $\comp$ uses $\bigo{1}$ time.
    \item Function \functionfont{BuildStack} runs in $\bigo{\ell\times\card{\Delta}}$ time (in fact in $\bigo{\ell\times(\card{Q}+\card{\Delta})}$).
    \item A call to function \functionfont{NextWord} runs in $\bigo{\ell\times\card{\Delta}}$ time.

    First, let us consider the operations that will be executed once.
    The creation of the empty state set $P$ uses $\bigo{\ell\times\card{\Delta}}$ time (see hypothesis \ref{hypothesis:state set}), which is in $\bigo{\ell\times\card{\Delta}}$.
    Note that each call to \functionfont{NextWord} gives rise to at most one call to \functionfont{MinWord} that returns a value other than~$\bot$, since \functionfont{NextWord} returns in that case.
    The cost of this call to \functionfont{MinWord} is in $\bigo{\ell}$, hence in $\bigo{\ell\times\card{\Delta}}$; the calls to \functionfont{MinWord} that return $\bot$ are treated below.

    Second, Let us show that the remainder of the code inside the \textbf{for} loop at line \ref{nextword:i loop start} is in $\bigo{\Delta}$.
    \begin{itemize}[align=left,leftmargin=*]
        \item
        Let us write~$P_a=\Delta(cur,a)$, this corresponds to line~\ref{nextword:delta}.
        To compute~$P_a$, we range over states in~$cur$ and browse the transitions labelled by $a$ and going out from states in~$cur$. Hence, computing $P_a$ uses $\bigo{\card{\Delta_a}}$ time,
        where $\Delta_a=\setst{(p,b,q)\in\Delta}{b=a}$; note also that $\card{P_a}\leq \card{\Delta_a}$.
        In the worst case, we compute $P_a$ for each letter $a$, which takes a total time in $\bigo{\sum_{a\in\Sigma}\card{\Delta_a}}=\bigo{\card{\Delta}}$
        \item
        The calls to \functionfont{MinWord} that return~$\bot$ costs $\bigo{\sum_{a\in\Sigma}\card{P_a}}$ time in total,
        which is in $\bigo{\card{\Delta}}$.
        \item  Clearing $P_a$ uses $\bigo{\card{P_a}}$ time (from hypothesis 
        \ref{hypothesis:state set}) and each $P_a$ is cleared at most once. Clearing all $P_a{}'s$ thus uses $\bigo{\card{\Delta}}$ time.
    \end{itemize}
\end{itemize}
The first output requires one call to \functionfont{MinWord}.
Each subsequent output requires one call to \functionfont{BuildStack} and one call to \functionfont{NextWord}.
Thus, the enumeration delay is in $\bigo{\ell\times\card{\Delta}}$ time, as claimed by  Theorem~\ref{t:ackerman-shallit}. 

\begin{remark}
   The enumeration phase of our algorithm is in some sense \textbf{memoryless}.
   Indeed, the preprocessed data structures ($\minarrow$ and $\comp$) are not modified, and no new persistent data structure is constructed.
   In other words, the $(i+1)$-th word to output is computed directly from the $i$-th word and the preprocessed data structures.
   This ensures in particular that the total space usage remains low. Indeed, recall that a polynomial delay algorithm could end up using exponential space after outputting an exponential number of results. This cannot happen with an algorithm with \textbf{memoryless polynomial delay} \cite{Strozecki2021} (sometimes also called \textbf{strong polynomial delay} \cite{Vigny2018}). 
\end{remark}

\section{Bibliography}
\printbibliography[heading=none, notkeyword={language}]

\end{document}

%% file: aux/vmthm.tex
\newlength{\thmspace}
\newtheoremstyle{vmstyle}%
    {\thmspace} 
    {\thmspace} 
    {\slshape\renewcommand{\emph}[1]{\textit{#1}}} 
    {} 
    {\bfseries} 
    {.} 
    {2ex} 
    {} 
\theoremstyle{vmstyle}

\newcommand{\thmBlockFont}[1]{#1}

\newcounter{thm}

\newtheorem{definition}[thm]{\thmBlockFont{Definition}}

\newtheorem{remark}[thm]{\thmBlockFont{Remark}}

\newtheorem{theorem}[thm]{\thmBlockFont{Theorem}}

\newtheorem{falsepropositionX}{\thmBlockFont{Proposition}}

\newtheorem{falsetheoremX}{\thmBlockFont{Theorem}}

\newtheorem{falsecorollaryX}{\thmBlockFont{Corollary}}

\newtheorem{falselemmaX}{\thmBlockFont{Lemma}}

\newtheorem*{falsestatementX}{\thmBlockFont{\thestatement}}

%% file: aux/tikzstyle.tex
\newlength{\minnoderadius}
\newlength{\shortenlength}
\newlength{\nodelinewidth}
\newlength{\arrowwidth}
\def\loopangle{24}
\tikzset{%
    bend angle=20,
    >={Stealth[width=6pt,length=6pt]},
    node/.style={circle, line width=\nodelinewidth, draw, black, inner sep=2pt, outer sep=.5*\nodelinewidth, minimum height=\minnoderadius, minimum width=\minnoderadius},
    vertex/.style={fill=black,inner sep=1.5pt, circle},
    state/.style={node},  
    run state/.style={draw,rounded rectangle},
    preedge/.style={-, draw, black, line width=1pt, rounded corners=5pt,pos=.4, shorten >=\shortenlength},
    edge/.style={preedge,->},
    redge/.style={preedge,<-},
    initialedge/.style={edge, shorten > =0pt, shorten < =0pt},
    finaledge/.style={redge, shorten > =0pt, shorten < =0pt},
    borderedge/.style={edge, -, color=white, line width=5pt, shorten >=\shortenlength-2pt, >={Stealth[width=12pt,length=12pt]}},
    vm loop/.style={edge, pos=.5, looseness = 8},    
    graph loop/.style={edge, pos=.5, looseness = 40, shorten >=4pt},    
    road/.style={edge, color=fred},
    hl/.style={edge, line width = 1.5mm, color=borange, shorten >=1pt},
    ferry/.style={edge,color=fpurple},
    gas/.style={edge,color=fblue},
    start/.style={edge,color=fblue},
    end/.style={edge,color=fblue},
    >=stealth,
    north west loop/.style={vm loop, in={\the\numexpr 135 + \loopangle\relax}, 
                                     out ={\the\numexpr 135 - \loopangle\relax}},
    north east loop/.style={vm loop, in={\the\numexpr 45 + \loopangle\relax}, 
                                     out ={\the\numexpr 45 - \loopangle\relax}},
    south west loop/.style={vm loop, in={\the\numexpr -135 + \loopangle\relax}, 
                                     out ={\the\numexpr -135 - \loopangle\relax}},
    south east loop/.style={vm loop, in={\the\numexpr -45 + \loopangle\relax}, 
                                     out ={\the\numexpr -45 - \loopangle\relax}},
    north loop/.style={vm loop, in={\the\numexpr 90 + \loopangle\relax}, 
                                out ={\the\numexpr 90 - \loopangle\relax}},
    south loop/.style={vm loop, in={\the\numexpr 270 - \loopangle\relax}, 
                                out ={\the\numexpr 270 + \loopangle\relax}},
    east loop/.style={vm loop, in={\the\numexpr 0 + \loopangle\relax}, 
                                out ={\the\numexpr 0 - \loopangle\relax}},
    west loop/.style={vm loop, in={\the\numexpr 180 - \loopangle\relax}, 
                                out ={\the\numexpr 180 + \loopangle\relax}},                
    node distance = \nodedist,
}

\newlength{\nodedist}
\newlength{\initfinaldist}
\newcommand{\initialfinal}[2][0]{%
    \def\angleI{\the\numexpr #1 + 15 \relax}
    \def\angleII{\the\numexpr #1 - 15 \relax}
    \path (#2.\angleI) ++(#1:\initfinaldist) coordinate 
        (#2-initialfinal1-#1);
    \path[initialedge] (#2.\angleI) to         
        (#2-initialfinal1-#1);
    \path (#2.\angleII) ++(#1:\initfinaldist) coordinate     (#2-initialfinal2-#1);
    \path[finaledge] (#2.\angleII) to
      (#2-initialfinal2-#1);
}
\definecolor{vert}{rgb}{0,.55,0.20}
\definecolor{bleu}{rgb}{0,0,0.75}
\definecolor{rouge}{rgb}{.75,0,0}